# A demonstration of the necessity of regularization in order to avoid inconsistent results in quantum field theory.

Dan Solomon
Rauland-Borg Corporation
1802 W. Central
Mount Prospect, IL
dan.solomon@rauland.com

January 1, 2013

**Abstract.**

We will examine a particular mathematical derivation in a paper by P. Falkensteiner and H. Grosse (F&G) [1]. In [1] a quantity $\Delta(A)$ is defined. This quantity is generated when the normal ordered generalized charge operator undergoes a unitary transformation. Using standard mathematical techniques F&G convert $\Delta(A)$ from its original form to another form which is suppose to be equivalent. It will be shown here that the two forms are not equivalent and that there is a mathematical inconsistency in their derivation. We will examine the source of this inconsistency and show that it can be resolved by proper regularization of the mathematical expressions.

## 1. Introduction.

In this paper we will examine the some issues regarding the problem of unitary implementation in quantum field theory. (Some work on this problem is given by these references [1-7].) Specifically we will examine a particular mathematical derivation in a paper that was published in 1988 by P. Falkensteiner and H. Grosse [1] (hereafter referred to as F&G). It will be shown that there is a mathematical inconsistency in one of the calculations in this paper.

F&G define a quantity $\Delta(A)$. From Eq. (4.12) of their paper it is shown that,

$$\Delta(A) \equiv TrP_-^0 VAV^\dagger P_-^0 - TrP_-^0 A P_-^0 \qquad (1.1)$$

The elements that make up this expression will be defined in more detail later but for the purposes of this section $P_-^0$ ($P_+^0$) is a projection operator on to the negative (positive) energy states, $A$ is a self-adjoint operator, and $V$ is a unitary operator.

In Eqs. (4.11) through (4.14) of their paper, F&G show that $\Delta(A)$ can be re-written as,

$$\Delta(A) = TrP_+^0 AP_- P_+^0 - TrP_-^0 AP_+ P_-^0 \tag{1.2}$$

where,

$$P_\pm = V^\dagger P_\pm^0 V \tag{1.3}$$

The question that we will address in this paper is whether or not these two expressions for $\Delta(A)$ are actually equivalent as claimed by F&G. In the next section we will pick a value of $A$ and $V$ so that $\Delta(A)$ as defined by (1.1) is trivially zero. However in Section 3 it will be shown that when $\Delta(A)$ is given by (1.2) it is non-zero. Therefore, we have an inconsistency. The reason why this inconsistency occurs will be examined in Section 4. It will be shown that inconsistency goes away if the integrals used to evaluate the above expressions are properly regularized.

## 2. A simple field theory.

F&G examine quantum field theory in (1+1) dimensions for zero mass fermions. For this case the free field Dirac equation is give by,

$$i\frac{\partial \psi}{\partial t} = -i\sigma_3 \frac{\partial \psi}{\partial x} \tag{2.1}$$

In order to simplify the discussion as much as possible we will not use the above equation but will develop a simpler field theory based on the equation,

$$i\frac{\partial \varphi}{\partial t} = -i\frac{\partial \varphi}{\partial x} \tag{2.2}$$

Define the Hamiltonian operator,

$$h_0 = -i\frac{\partial}{\partial x} \tag{2.3}$$

The solutions to (2.2) are $\varphi_q^\pm(x,t) = e^{\mp iqt}\phi_q^\pm(x)$ with $q \geq 0$ where $\phi_q^\pm(x)$ are solutions to,

$$h_0\phi_q^\pm(x) = \pm q\phi_q^\pm(x) \tag{2.4}$$

and are given by,

$$\phi_q^\pm(x) = \frac{e^{\pm iqx}}{\sqrt{2\pi}} \quad (2.5)$$

The $\phi_q^+(x)$ are the positive energy solutions and $\phi_q^-(x)$ are the negative energy solutions. From this we define the positive and negative energy projection operators, $P_+^0$ and $P_-^0$, respectively as,

$$P_\mp^0 f(x) = \frac{1}{2\pi} \int_0^\infty dk\, e^{\mp ikx} \langle e^{\pm iky} f(y) \rangle \quad (2.6)$$

where $\langle g(y) \rangle \equiv \int g(y) dy$. The projection operators obey,

$$P_\pm^0 P_\pm^0 = P_\pm^0,\ P_\pm^0 P_\mp^0 = 0,\ \text{and}\ P_+^0 + P_-^0 = 1 \quad (2.7)$$

The above relationships are also satisfied by $P_\pm$ which was defined above by Eq. (1.3).

Following F&G define the field operator $\hat{\psi}$ according to,

$$\hat{\psi}(f) = \hat{b}(P_+^0 f) + \hat{d}^\dagger\left((P_-^0 f)^*\right) \quad (2.8)$$

where the operators $\hat{b}$ and $\hat{d}$ obey the usual canonical anticommutation relations (CAR),

$$\{\hat{b}(P_+^0 f), \hat{b}^\dagger(P_+^0 g)\} = \langle f, P_+^0 g \rangle, \qquad \{\hat{d}(P_-^0 f), \hat{d}^\dagger(P_-^0 g)\} = \langle f, P_-^0 g \rangle \quad (2.9)$$

Define the generalized charge operator,

$$q(A) = \sum_{p,q>0} \left( b_p^\dagger A_{pq}^{++} b_q + b_p^\dagger A_{pq}^{+-} d_q^\dagger + d_p A_{pq}^{-+} b_q + d_p A_{pq}^{--} d_q^\dagger \right) \quad (2.10)$$

which holds for trace class operators A where,

$$A_{pq}^{\pm\pm} = \langle \phi_p^\pm, A\phi_q^\pm \rangle,\ A_{pq}^{\pm\mp} = \langle \phi_p^\pm, A\phi_q^\mp \rangle,\ b_p = b(\phi_p^+),\ d_p = d(\phi_p^-) \quad (2.11)$$

Next F&G introduce $Q(A)$ as the normal ordering of the operator $q(A)$ to obtain,

$$Q(A) = \sum_{p,q>0} \left( b_p^\dagger A_{pq}^{++} b_q + b_p^\dagger A_{pq}^{+-} d_q^\dagger + d_p A_{pq}^{-+} b_q - d_q^\dagger A_{pq}^{--} d_p \right)$$
$$= q(A) - Tr P_-^{(0)} A P_-^{(0)} \quad (2.12)$$

According to [8] the trace class restriction on $A$ can be relaxed for the operator $Q(A)$. Assume that the unitary operator $V$ is implemental. This means that there exists a $\Gamma(V)$ where,

$$\psi(Vf) = \Gamma(V)\psi(f)\Gamma^\dagger(V) \tag{2.13}$$

F&G apply this unitary transformation to $Q(A)$ to obtain,

$$\Gamma(V)Q(A)\Gamma^\dagger(V) = Q(VAV^\dagger) + \Delta(A) \tag{2.14}$$

where $\Delta(A)$ is given by (1.1) and can be written as,

$$\Delta(A) = F_1(A) - F_2(A) \tag{2.15}$$

with,

$$F_1(A) = TrP_-^{(0)} VAV^\dagger P_-^{(0)} \tag{2.16}$$

and,

$$F_2(A) = TrP_-^{(0)} A P_-^{(0)} \tag{2.17}$$

The Trace is given by,

$$TrG(x,y) = \frac{1}{2\pi} \int_{-\infty}^{+\infty} dp \iint e^{ipx} G(x,y) e^{-ipy} dxdy \tag{2.18}$$

Let the operator $A$ be a real valued function $A(x)$ for which $A(\pm\infty) \to 0$ and which satisfies the condition,

$$\int A(x)dx = 0 \tag{2.19}$$

and let the unitary operator $V$ be given by,

$$V(x) = e^{iC(x)}, \quad V^\dagger(x) = e^{-iC(x)} \tag{2.20}$$

where $C(x)$ is a real valued function with $C(\pm\infty) \to 0$. Based on the discussion in Refs [6,7] it can shown that $V(x)$ is unitary implemental.

Evaluate (2.17) and use (2.19) to obtain,

$$F_2(A) = \frac{1}{2\pi} \int_0^\infty dp \int \left(e^{+ipx} A(x) e^{-ipx}\right) dx = \frac{1}{2\pi} \int_0^\infty dp \int A(x) dx = 0 \tag{2.21}$$

Next, use (2.20) to obtain $VAV^\dagger = A$. Use this in (2.16) to yield,

$$F_1(A) = F_2(A) \tag{2.22}$$

Use this and (2.21) in (2.15) to obtain,

$$\Delta(A) = 0 \tag{2.23}$$

Referring to (2.16) and using $VV^\dagger = 1$ along with (1.3) and the cyclic property of the trace we can rewrite $F_1(A)$ as,

$$F_1(A) = TrVV^\dagger P_-^{(0)} VAV^\dagger P_-^{(0)} = TrP_- AP_- = TrAP_- \quad (2.24)$$

Next use the relationship $P_-^{(0)} + P_+^{(0)} = 1$ to obtain

$$F_1(A) = Tr\left(P_-^{(0)} + P_+^{(0)}\right) AP_- = F_{1a}(A) + F_{1b}(A) \quad (2.25)$$

where,

$$F_{1a}(A) = TrP_-^{(0)} AP_- \text{ and } F_{1b}(A) = TrP_+^{(0)} AP_- \quad (2.26)$$

Next, referring to (2.17) we use the cyclic property of the trace to obtain,

$$F_2(A) = TrP_-^{(0)} A \quad (2.27)$$

Use $P_- + P_+ = 1$ in the above to obtain,

$$F_2(A) = TrP_-^{(0)} A(P_- + P_+) = F_{2a}(A) + F_{2b}(A) \quad (2.28)$$

where,

$$F_{2a}(A) = TrP_-^{(0)} AP_- \text{ and } F_{2b}(A) = TrP_-^{(0)} AP_+ \quad (2.29)$$

From the above relationships we can write,

$$\Delta(A) = \Delta'(A) + \left(F_{1a}(A) - F_{2a}(A)\right) \quad (2.30)$$

where,

$$\Delta'(A) \equiv F_{1b}(A) - F_{2b}(A) \quad (2.31)$$

Note that $F_{1a}(A) = F_{2a}(A)$ since the expressions are identical (see (2.26) and (2.29)). Therefore $\Delta'(A) = \Delta(A)$ which, from (2.23), means that $\Delta'(A) = 0$. However as will be shown in the next section when we calculate $\Delta'(A)$ as defined by (2.31) the result is in general non-zero. Therefore we have an inconsistency.

### 3. Demonstrating an inconsistency.

In this section we will evaluate $\Delta'(A)$ as defined by (2.31). Evaluate $F_{1b}(A)$ as defined by Eq. (2.26) to obtain,

$$F_{1b}(A) = \frac{1}{4\pi^2} \int_0^\infty dp \int_0^\infty dk \left\langle e^{-i(p+k)x} A(x) V^\dagger(x) \right\rangle \left\langle V(y) e^{+i(p+k)y} \right\rangle \quad (3.1)$$

Perform a change of variables, $s = p+k$ and $r = p-k$ so that,

$$F_{1b}(A) = \frac{1}{8\pi^2} \int_0^{+\infty} ds \int_{-s}^{+s} dr \left\langle e^{-isx} A(x) V^\dagger(x) \right\rangle \left\langle V(y) e^{+isy} \right\rangle$$

$$= \frac{1}{4\pi^2} \int_0^{+\infty} sds \left\langle e^{-isx} A(x) V^\dagger(x) \right\rangle \left\langle V(y) e^{+isy} \right\rangle \quad (3.2)$$

This can be written as,

$$F_{1b}(A) = -\frac{i}{4\pi^2} \int_0^{+\infty} ds \left\langle e^{-isx} A(x) V^\dagger(x) \right\rangle \left\langle V(y) \frac{d}{dy} e^{+isy} \right\rangle$$

$$= \frac{i}{4\pi^2} \int_0^{+\infty} ds \left\langle e^{-isx} A(x) V^\dagger(x) \right\rangle \left\langle e^{+isy} \frac{d}{dy} V(y) \right\rangle \quad (3.3)$$

where we have used integration by parts to obtain the last result. By a similar analysis refer to (2.29) obtain,

$$F_{2b}(A) = -\frac{i}{4\pi^2} \int_0^{\infty} ds \left\langle e^{isx} A(x) V^\dagger(x) \right\rangle \left\langle e^{-isy} \frac{d}{dy} V(y) \right\rangle \quad (3.4)$$

Use the above results in (2.31) to yield,

$$\Delta'(A) = \frac{i}{4\pi^2} \iint dxdy A(x) V^\dagger(x) \frac{dV(y)}{dy} \int_{-\infty}^{+\infty} e^{is(x-y)} ds \quad (3.5)$$

Use $\int_{-\infty}^{+\infty} e^{is(x-y)} ds = 2\pi\delta(x-y)$ in the above to obtain,

$$\Delta'(A) = \frac{i}{2\pi} \int A(x) V^\dagger(x) \frac{dV(x)}{dx} dx \quad (3.6)$$

This result is consistent with Eq. (4.19) of F&G. In general this expression is non-zero. This is in contradiction to our previous result were it was shown that $\Delta'(A) = \Delta(A) = 0$.

## 4. Regularizing the integrals.

To review the results of the previous section we start with the quantity $\Delta(A)$ which is given by (1.1). We pick $A$ and $V$ so that $\Delta(A)$ is zero. Next we use a straightforward mathematic procedure to obtain the quantity $\Delta'(A)$ which is given by (2.31). It is shown that $\Delta(A) = \Delta'(A)$ therefore we could write $\Delta(A) = F_{1b}(A) - F_{2b}(A)$ which is equivalent to Eq. (1.2) which is the alternative expression for $\Delta(A)$ that is given by F&G. However in the last section we have shown that $\Delta'(A)$ is given by (3.6) and is, in general, non-zero. Therefore the results are inconsistent.

In order to try to resolve this problem evaluate $F_{1a}(A)$ as defined in (2.26) to obtain,

$$F_{1a}(A) = \frac{1}{4\pi^2}\int_0^\infty dp \int_0^\infty dk \left\langle e^{i(p-k)x} A(x) V^\dagger(x) \right\rangle \left\langle V(y) e^{-i(p-k)y} \right\rangle \tag{4.1}$$

We modify this expression slightly as follows,

$$F_{1a}(A;\varepsilon_1,\varepsilon_2) = \frac{1}{4\pi^2}\int_0^\infty e^{-p\varepsilon_1} dp \int_0^\infty e^{-k\varepsilon_2} dk \left\langle e^{i(p-k)x} A(x) V^\dagger(x) \right\rangle \left\langle V(y) e^{-i(p-k)y} \right\rangle \tag{4.2}$$

In the above expression we have "regularized" the integrals by multiplying by the factors $e^{-p\varepsilon_1}$ and $e^{-k\varepsilon_2}$ where $\varepsilon_1, \varepsilon_2 \to 0$ and $\varepsilon_1, \varepsilon_2 > 0$. Similarly, for $F_{1b}(A)$ we write,

$$F_{1b}(A;\varepsilon_1,\varepsilon_2) = \frac{1}{4\pi^2}\int_0^\infty e^{-p\varepsilon_1} dp \int_0^\infty e^{-k\varepsilon_2} dk \left\langle e^{-i(p+k)x} A(x) V^\dagger(x) \right\rangle \left\langle V(y) e^{+i(p+k)y} \right\rangle \tag{4.3}$$

Perform the integrations with respect $p$ and $k$ to obtain,

$$F_{1a}(A;\varepsilon_1,\varepsilon_2) = \frac{1}{4\pi^2}\int dx A(x) V^\dagger(x) \int dy V(y) \left(\frac{1}{(y-x)-i\varepsilon_1}\right)\left(\frac{1}{(y-x)+i\varepsilon_2}\right) \tag{4.4}$$

$$F_{1b}(A;\varepsilon_1,\varepsilon_2) = -\frac{1}{4\pi^2}\int dx A(x) V^\dagger(x) \int dy V(y) \left(\frac{1}{(y-x)+i\varepsilon_1}\right)\left(\frac{1}{(y-x)+i\varepsilon_2}\right) \tag{4.5}$$

Use these two equations in (2.25) to obtain,

$$F_1(A;\varepsilon_1,\varepsilon_2) = \frac{i}{2\pi}\int dx \int dy \delta(y-x;\varepsilon_1) \left(\frac{1}{(y-x)+i\varepsilon_2}\right) A(x) V^\dagger(x) V(y) \tag{4.6}$$

where,

$$\delta(w;\varepsilon) \equiv \frac{1}{\pi}\frac{\varepsilon}{w^2+\varepsilon^2} \tag{4.7}$$

It is shown in the Appendix that this becomes,

$$F_1(A;\varepsilon_1,\varepsilon_2) = i\frac{J(\varepsilon_1,\varepsilon_2)}{2\pi}\int dx A(x) V^\dagger(x) \frac{dV(x)}{dx} \tag{4.8}$$

where,

$$J(\varepsilon_1, \varepsilon_2) = \frac{\varepsilon_1}{(\varepsilon_1 + \varepsilon_2)} \qquad (4.9)$$

According to Eq. (2.22), $F_1(A) = 0$. It is evident that $F_1(A; \varepsilon_1, \varepsilon_2)$ is not automatically zero even as $\varepsilon_1, \varepsilon_2 \to 0$ but depends on how these two quantities go to zero. For example if $\varepsilon_2 = \varepsilon_1 \to 0$ then $J(\varepsilon_1, \varepsilon_2) = 1/2$ and $F_1(A; \varepsilon_1, \varepsilon_2)$ will not be zero. In order to ensure that $F_1(A; \varepsilon_1, \varepsilon_2)$ goes to zero we must select $\varepsilon_1$ and $\varepsilon_2$ so that they approach zero in such a way that they force $J(\varepsilon_1, \varepsilon_2)$ to approach zero. The condition is that as $\varepsilon_1$ and $\varepsilon_2$ approach zero the ratio $\varepsilon_1/\varepsilon_2 \to 0$. This can be assured by setting,

$$\varepsilon_2 = \sqrt{\varepsilon_1} \qquad (4.10)$$

In this case,

$$J(\varepsilon_1, \varepsilon_2) = \varepsilon_1 / (\varepsilon_1 + \sqrt{\varepsilon_1})\Big|_{\varepsilon_1 \to 0} = \sqrt{\varepsilon_1}\Big|_{\varepsilon_1 \to 0} = 0 \qquad (4.11)$$

Use the same approach to evaluate $F_{2a}(A)$ and $F_{2b}(A)$ to obtain,

$$F_{2a}(A; \varepsilon_1, \varepsilon_2) = \frac{1}{4\pi^2} \int_0^{+\infty} dp\, e^{-p\varepsilon_2} \int_0^{+\infty} dk\, e^{-k\varepsilon_1} \left\langle e^{i(p-k)x} A(x) V^\dagger(x) \right\rangle \left\langle V(y) e^{-i(p-k)y} \right\rangle \qquad (4.12)$$

$$F_{2b}(A; \varepsilon_1, \varepsilon_2) = \frac{1}{4\pi^2} \int_0^{+\infty} dp\, e^{-p\varepsilon_2} \int_0^{+\infty} dk\, e^{-k\varepsilon_1} \left\langle e^{i(p+k)x} A(x) V^\dagger(x) \right\rangle \left\langle V(y) e^{-i(p+k)y} \right\rangle \qquad (4.13)$$

Solve the above to obtain,

$$F_{2a}(A; \varepsilon_1, \varepsilon_2) = \frac{1}{4\pi^2} \int dx A(x) V^\dagger(x) \int dy V(y) \left(\frac{1}{(y-x)-i\varepsilon_2}\right)\left(\frac{1}{(y-x)+i\varepsilon_1}\right) \qquad (4.14)$$

$$F_{2b}(A; \varepsilon_1, \varepsilon_2) = -\frac{1}{4\pi^2} \int dx A(x) V^\dagger(x) \int dy V(y) \left(\frac{1}{(y-x)-i\varepsilon_2}\right)\left(\frac{1}{(y-x)-i\varepsilon_1}\right) \qquad (4.15)$$

Use the above in Eq. (2.28) to obtain,

$$F_2(A; \varepsilon_1, \varepsilon_2) = \frac{-i}{2\pi} \int dx \int dy A(x) V^\dagger(x) V(y) \delta(y-x; \varepsilon_1) \left(\frac{1}{(y-x)-i\varepsilon_2}\right) \qquad (4.16)$$

Using steps similar to the evaluation of $F_1(A; \varepsilon_1, \varepsilon_2)$ we obtain,

$$F_2(A; \varepsilon_1, \varepsilon_2) = -i \frac{J(\varepsilon_1, \varepsilon_2)}{2\pi} \int dx A(x) V^\dagger(x) \frac{dV(x)}{dx} \qquad (4.17)$$

As before set $\varepsilon_2 = \sqrt{\varepsilon_1}$ which will make $F_2(A;\varepsilon_1,\varepsilon_2) \to 0$ as $\varepsilon_1 \to 0$.

Use the above in Eq. (2.31) to obtain,

$$\Delta'(A;\varepsilon_1,\varepsilon_2) = F_{1b}(A;\varepsilon_1,\varepsilon_2) - F_{2b}(A;\varepsilon_1,\varepsilon_2) \tag{4.18}$$

From the above discussion we obtain,

$$\Delta'(A;\varepsilon_1,\varepsilon_2) = -\frac{1}{4\pi^2}\int dx A(x)V^\dagger(x)\int dy V(y) \left\{ \frac{1}{[(y-x)+i\varepsilon_1][(y-x)+i\varepsilon_2]} - \frac{1}{[(y-x)-i\varepsilon_2][(y-x)-i\varepsilon_1]} \right\} \tag{4.19}$$

In the Appendix this is shown to be,

$$\Delta'(A;\varepsilon_1,\varepsilon_2) = \frac{i}{2\pi}\int dx A(x)V^\dagger(x)\frac{dV(x)}{dx} \tag{4.20}$$

This is the same result that was achieved in the last section for $\Delta'(A)$ (see Eq. (3.6)). The key difference between the results of this section and the last section has to do with the evaluation of the quantity $F_{1a}(A) - F_{2a}(A)$. $\Delta(A)$ and $\Delta'(A)$ are related by,

$$\Delta(A) = \Delta'(A) + (F_{1a}(A) - F_{2a}(A)) \tag{4.21}$$

In the last section it was shown that $F_{1a}(A) - F_{2a}(A) = 0$. However using the results of this section we obtain,

$$F_{1a}(A;\varepsilon_1,\varepsilon_2) - F_{2a}(A;\varepsilon_1,\varepsilon_2) = \int dx A(x)V^\dagger(x)\int dy V(y)\left(\frac{1}{[(y-x)-i\varepsilon_1][(y-x)+i\varepsilon_2]} - \frac{1}{[(y-x)-i\varepsilon_2][(y-x)+i\varepsilon_1]}\right)$$

$$\tag{4.22}$$

This quantity will be zero if $\varepsilon_1 = \varepsilon_2$. However, as has already been discussed, $\varepsilon_2 \ne \varepsilon_1$. In the Appendix this is shown to be,

$$F_{1a}(A;\varepsilon_1,\varepsilon_2) - F_{2a}(A;\varepsilon_1,\varepsilon_2) = \frac{i}{2\pi}\frac{(\varepsilon_1-\varepsilon_2)}{(\varepsilon_1+\varepsilon_2)}\int dx A(x)V^\dagger(x)\frac{dV(x)}{dx} \tag{4.23}$$

Use this result along with (4.20) and (4.21) to obtain,

$$\Delta(A;\varepsilon_1,\varepsilon_2) = \frac{iJ(\varepsilon_1,\varepsilon_2)}{\pi}\int dx A(x)V^\dagger(x)\frac{dV(x)}{dx} \tag{4.24}$$

As discussed above if $\varepsilon_1$ and $\varepsilon_2$ go to zero while maintaining the relationship $\varepsilon_2 = \sqrt{\varepsilon_1}$ then $J(\varepsilon_1, \varepsilon_2) \to 0$ which means that $\Delta(A; \varepsilon_1, \varepsilon_2) \to 0$.

## 5. Conclusion.

We have examined the quantity $\Delta(A)$ which is defined by Eq. (1.1). Using formal mathematical techniques it can be converted into the form given by Eq. (1.2). However when these two expressions are evaluated they are found to yield different results. The reason why this problem occurs is made apparent when we regularized the integrations in the various expressions by including factors of the form $e^{-p\varepsilon}$ and letting $\varepsilon \to 0$. Since there are two integrations that go to infinity there are two factors and two infinitesimal terms $\varepsilon_1$ and $\varepsilon_2$. As discussed above proper regularization requires that $\varepsilon_1$ goes to zero much faster than $\varepsilon_2$. This is guaranteed by requiring that $\varepsilon_2 = \sqrt{\varepsilon_1}$. When this is done both expressions for $\Delta(A)$ yield the same result for the problem we have considered and the inconsistency is resolved.

## Appendix

We will calculate $F_1(A; \varepsilon_1, \varepsilon_2)$ as defined by (4.6) which we rewrite below for convience.

$$F_1(A; \varepsilon_1, \varepsilon_2) = \frac{i}{2\pi} \int dx \int dy \delta(y - x; \varepsilon_1) \left( \frac{1}{(y-x) + i\varepsilon_2} \right) A(x) V^\dagger(x) V(y) \quad (6.1)$$

The integrand is small, on the order of $\varepsilon_1 \to 0$, except when $y \to x$. Therefore we can expand $V(y)$ as follows,

$$V(y) \cong V(x) + \frac{dV(x)}{dx}(y - x) \quad (6.2)$$

Use this in (6.1) to obtain,

$$F_1(A) = D_1(A) + D_2(A) \quad (6.3)$$

where,

$$D_1(A) = \frac{i}{2\pi^2} \int dx A(x) \int dy \left( \frac{\varepsilon_1}{(y-x)^2 + \varepsilon_1^2} \right) \left( \frac{1}{(y-x) + i\varepsilon_2} \right) \quad (6.4)$$

and,

$$D_2(A) = \frac{i}{2\pi^2} \int dx A(x) V^\dagger(x) \frac{dV(x)}{dx} \int dy \left( \frac{(y-x)\varepsilon_1}{(y-x)^2 + \varepsilon_1^2} \right) \left( \frac{1}{(y-x) + i\varepsilon_2} \right) \quad (6.5)$$

Performing the integration with respect to $y$ and using (2.19) we obtain,

$$D_1(A) = \left( \frac{1}{2\pi(\varepsilon_1 + \varepsilon_2)} \right) \int dx A(x) = 0 \quad (6.6)$$

and,

$$D_2(A) = \left( \frac{i\varepsilon_1}{2\pi(\varepsilon_1 + \varepsilon_2)} \right) \int dx A(x) V^\dagger(x) \frac{dV(x)}{dx} \quad (6.7)$$

Use this in (6.3) to obtain (4.8) in the text.

Next we evaluate Eq. (4.19) which is also written below,

$$\Delta'(A; \varepsilon_1, \varepsilon_2) = -\frac{1}{4\pi^2} \int dx A(x) V^\dagger(x) \int dy V(y) \left\{ \begin{array}{c} \dfrac{1}{[(y-x) + i\varepsilon_1][(y-x) + i\varepsilon_2]} \\ -\dfrac{1}{[(y-x) - i\varepsilon_2][(y-x) - i\varepsilon_1]} \end{array} \right\} \quad (6.8)$$

The integrand is small except when $y \to x$ therefore we can expand $V(y)$ according to Eq. (6.2) and use $V^\dagger V = 1$ to obtain,

$$\Delta'(A) = M_1(A) + M_2(A) \quad (6.9)$$

$$M_1(A) = \frac{-1}{4\pi^2} \int dx A(x) \int dy \left\{ \begin{array}{c} \dfrac{1}{[(y-x) + i\varepsilon_1][(y-x) + i\varepsilon_2]} \\ -\dfrac{1}{[(y-x) - i\varepsilon_2][(y-x) - i\varepsilon_1]} \end{array} \right\} \quad (6.10)$$

and,

$$M_2(A) = -\frac{1}{4\pi^2} \int dx A(x) V^\dagger(x) \frac{dV(x)}{dx} \int dy \left\{ \begin{array}{c} \dfrac{(y-x)}{[(y-x) + i\varepsilon_1][(y-x) + i\varepsilon_2]} \\ -\dfrac{(y-x)}{[(y-x) - i\varepsilon_2][(y-x) - i\varepsilon_1]} \end{array} \right\} \quad (6.11)$$

Using (2.19) will result in $M_1(A) = 0$. In the above equation perform the integration with respect to $y$ to obtain,

$$M_2(A) = \frac{i}{2\pi} \int dx A(x) V^\dagger(x) \frac{dV(x)}{dx} \tag{6.12}$$

Use these results in (6.9) to yields (4.20) in the text.

Next evaluate $F_{1a}(A; \varepsilon_1, \varepsilon_2) - F_{2a}(A; \varepsilon_1, \varepsilon_2)$ which is given by Eq. (4.22). As in the previous cases we can expand $V(y)$ according to (6.2) to obtain,

$$F_{1a}(A) - F_{2a}(A) = N_1(A) + N_2(A) \tag{6.13}$$

where,

$$N_1(A) = \frac{1}{4\pi^2} \int dx A(x) \int dy \left( \frac{1}{[(y-x)-i\varepsilon_1][(y-x)+i\varepsilon_2]} - \frac{1}{[(y-x)-i\varepsilon_2][(y-x)+i\varepsilon_1]} \right) \tag{6.14}$$

and,

$$N_2(A) = \frac{1}{4\pi^2} \int dx A(x) V^\dagger(x) \frac{dV(x)}{dx} \int dy \left( \frac{(y-x)}{[(y-x)-i\varepsilon_1][(y-x)+i\varepsilon_2]} - \frac{(y-x)}{[(y-x)-i\varepsilon_2][(y-x)+i\varepsilon_1]} \right) \tag{6.15}$$

Using (2.19),

$$N_1(A) = 0 \tag{6.16}$$

$N_2(A)$ can be shown to be,

$$N_2(A) = \frac{i}{2\pi} \frac{(\varepsilon_1 - \varepsilon_2)}{(\varepsilon_1 + \varepsilon_2)} \int dx A(x) V^\dagger(x) \frac{dV(x)}{dx} \tag{6.17}$$

Therefore,

$$F_{1a}(A) - F_{2a}(A) = \frac{i}{2\pi} \frac{(\varepsilon_1 - \varepsilon_2)}{(\varepsilon_1 + \varepsilon_2)} \int dx A(x) V^\dagger(x) \frac{dV(x)}{dx} \tag{6.18}$$

# References

1. P. Falkensteiner and H. Grosse. "Fermions in interaction with time-dependent fields". Nuclear Physics, B305 (1988) 126-142.

2. P. Falkensteiner and H. Grosse. "Quantization of Fermions Interacting with Point-like External Fields". Lett. Math. Phys.,**14** (1987) 139-148.

3. P. Falkensteiner and H. Grosse. "Unitary Implementability of gauge transformations for the Dirac operator and the Schwinger term" J. Math. Phys. **28** (4) 850-854.

4. S.N.M. Ruijsenaars. Gauge particles in external field. 1. Classical theory. J. Math. Phys. **18** (1977) 720-732.

5. S.N.M. Ruijsenaars. On Bogoliubov transformations for systems of relativistic charge particles. J. Math. Phys. **18** (1977) 517-526.

6. A.L. Carey, C.A. Hurst, and D.M. O'Brien. "Fermion Currents in 1+1 dimensions". J. Math. Phys. **24** (1983) 2212-2221.

7. A.L. Carey, C.A. Hurst, and D.M. O'Brien. "Automorphisms of the Canonical Anticommutation Relations and Index Theory". J. Funct. Anal. **48** (1982) 360-393.

8. Bernd Thaller. *The Dirac Equation.* Springer-Verlag, New York (1992).